\documentclass[a4paper]{spie}
\usepackage[left=1.925cm, right=1.925cm, top=2.54cm, bottom=4.94cm]{geometry}
\usepackage{graphicx,subfig}
\begin{document}

\title{Real-time measurement with fiber optical surface plasmon resonance sensor for biochemical interaction analysis}
\author{Zhixin Tan$^a$ and Xuejin Li$^{*a}$ and  Yuzhi Chen$^a$ and Ping Fan$^a$
	\skiplinehalf
	\supit{a}College of Physics Science and Technology, Shenzhen Key 
   Laboratory of Sensor Technology, Shenzhen University, Guangdong, 518060, China
}

\authorinfo{Corresponding author: lixuejin@szu.edu.cn}

\maketitle{}

\abstract{ In this paper we report a fiber optical sensor system based
  on surface plasmon resonance (SPR) with real-time response for
  biochemical interaction analysis. The fiber sensor is produce from a
  multi-mode fiber with plastic cladding. To facilitate the
  measurement, a software program is developed which integrates the
  data acquisition and processing for real-time feedback. Polynomial
  fitting is implemented to smooth out the noise in transmission ratio
  and a spectral resolution of 0.2 nm is achieved. Ethyl alcohol and
  water mixtures of different concentration are measured to
  demonstrate the system real-time ability. This work is essential for
  the development of a compact, real-time fiber SPR biosensor.

   }

\section{Introduction}

Fiber optical surface plasmon resonance (SPR) sensor was invented by
R.C. Jorgenson and S.S. Yee in 1983 as a novel chemical
sensor~\cite{Jorgenson1993fiberopticchemical}. Currently, with the
need for minimization and automation of detection in bio-medical
applications, fiber optical SPR sensors have attracted much interest
for its compactness and the flexibility. Fiber optical SPR sensor has
many features, including label-free detection, high selectivity, low
cost and immunity to electromagnetic interference. Since fiber SPR
sensor can conduct the measurement \emph{in situ} and process data
remotely, it may be useful in many potential
applications~\cite{Lee2009Currentstatusmicro,Rajan2007Surfaceplasmonresonance,Delport2012Realtimemonitoring,Verma2013FiberopticSPR},
such as environmental monitoring and medical detection. To achieve
this, real-time response is necessary for a measurement that feedback
the mass change on the surface instantly in a biochemical
interaction. In this paper, we describe our work toward a compact,
real-time response fiber SPR measure system.

\section{Principle}

\begin{figure}[htb]
  \centering
  \includegraphics[width=0.6\textwidth]{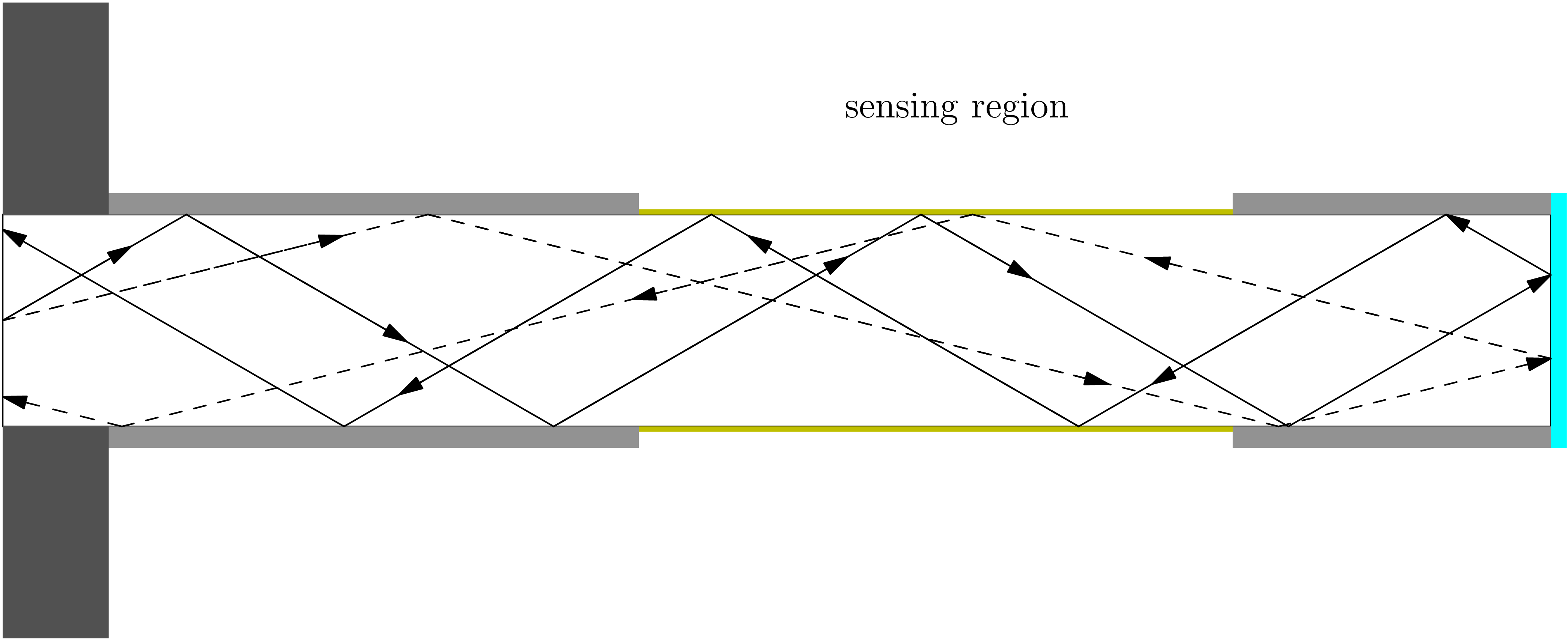}
  \caption{The principle of multi-mode fiber SPR sensor. The real and
    dashed lines depict the track of light ray of different incident
    angle.}
  \label{fig:mmfiber}
\end{figure}

Surface plasmon resonance measurement utilizes an evanescent field to
sense the change of the environmental refractive index. Meanwhile, an
optical fiber confines light by the total internal reflection, and the
evanescent field is along the interface of the fiber core and
cladding. Apparently, integrating the surface plasmon resonance region
along the fiber cylinder surface will make a compact probe which is
useful in medical and biological research. The principle of multi-mode
fiber optical SPR sensor is shown in Fig.\ref{fig:mmfiber}, the real
and dashed lines depict the track of light ray of different incident
angle. Incident light with angle greater than critical angle will be
confined inside the cylinder. Since the noble metal film is coating on
the cylindrical surface, the free electron will be driver by the
evanescent field in the plane and conduct a collective oscillation
which is a plasmonic wave.  Whenever the wave vector of the incident
light matches that of the surface plasmon, energy will transport
from input light to the plasmonic wave. An absorption dip will show in
the calculated transmission ratio. This dip is sensitive to the
change of refractive index that may associate with the mass change on
the fiber surface. This mass change may be a result of affinity
reaction.

Fiber SPR sensor may operate in three modes: angle interrogation,
amplitude interrogation, and wavelength interrogation. Angle
interrogation is derived from prism configuration by change the
incident angle of laser light. However, it is hard to keep the fiber
straight and confine the light at desired angle. Particularly, this
will fail under a remote measurement setting. Amplitude interrogation
will measure the fiber loss with light of single wavelength. Whenever
the environmental refractive index changes, the loss will decrease or
increase which is easy to measure. However, it will lose much
information about the resonance, such as whether the loss change is
from a surface plasmon resonance. We preferred the wavelength
interrogation which will precisely measure the resonance wavelength
with a fiber spectrometer. The theoretical model of the fiber optical
SPR sensor has been studied in
detail~\cite{Sharma2007FiberOpticSensors,Gupta2009Surfaceplasmonresonance,Yuan2011NumericalinvestigationSPR}. Here
we narrate the calculation shortly.
\begin{equation}
  P_{trans} = \frac{\int_{\theta_{cr}} ^{\pi/2}  (R_p^{N(\theta)} +
    R_s^{N(\theta)} ) P(\theta)  \mathrm{d}\theta }
   { 2\int_{\theta_{cr}}^{\pi/2} P(\theta)   \mathrm{d}\theta  }
   \label{eq:ptrans}
\end{equation}
where $ N(\theta) = D /tan{\theta} $, $\theta_{cr} = \sin^{-1}(n_{cl}
/n_1) $, $ P(\theta) = n_1^2 \sin\theta \cos\theta / (1-n_1^2
\cos^2\theta )^2 $, $N(\theta)$ is the number of reflections performed
by the ray with incident angle $\theta$, while $L$ and $D$ represent
the length of the sensing region and the diameter of fiber core;
$\theta_{cr}$ is the critical angle of the fiber whereas $n_{cl}$ and
$n_1$ are the refractive index of the fiber cladding and
core. $P_\theta$ presents the modal power as a function of the
incident angle $\theta$. $R_p$ and $R_s$ are the reflection ratio of
p- or s- polarized ray.

In our configuration we don't distinguish the light in p- or s-
polarization. Therefore the light of s- polarization is included in
the spectra which will pad the curves. According to
Eq.~\ref{eq:ptrans}, the curve of the reflection ratio is not a valley
with symmetric shape, but a result of mixed system with broadband
light ray in various incident angles. This theoretical model
establishes a basement for the data fitting scheme in our software
development.

\section{System implementation and integration}
Here we show our experimental configuration and the measurement system which is dedicated for the fiber optical SPR sensor. Then we discuss the data fitting scheme and the real-time performance of the fiber sensor system.

\begin{figure}[tbh]
  \centering
  \subfloat[]{
    \label{fig:schematic}
    \includegraphics[width=0.4\textwidth]{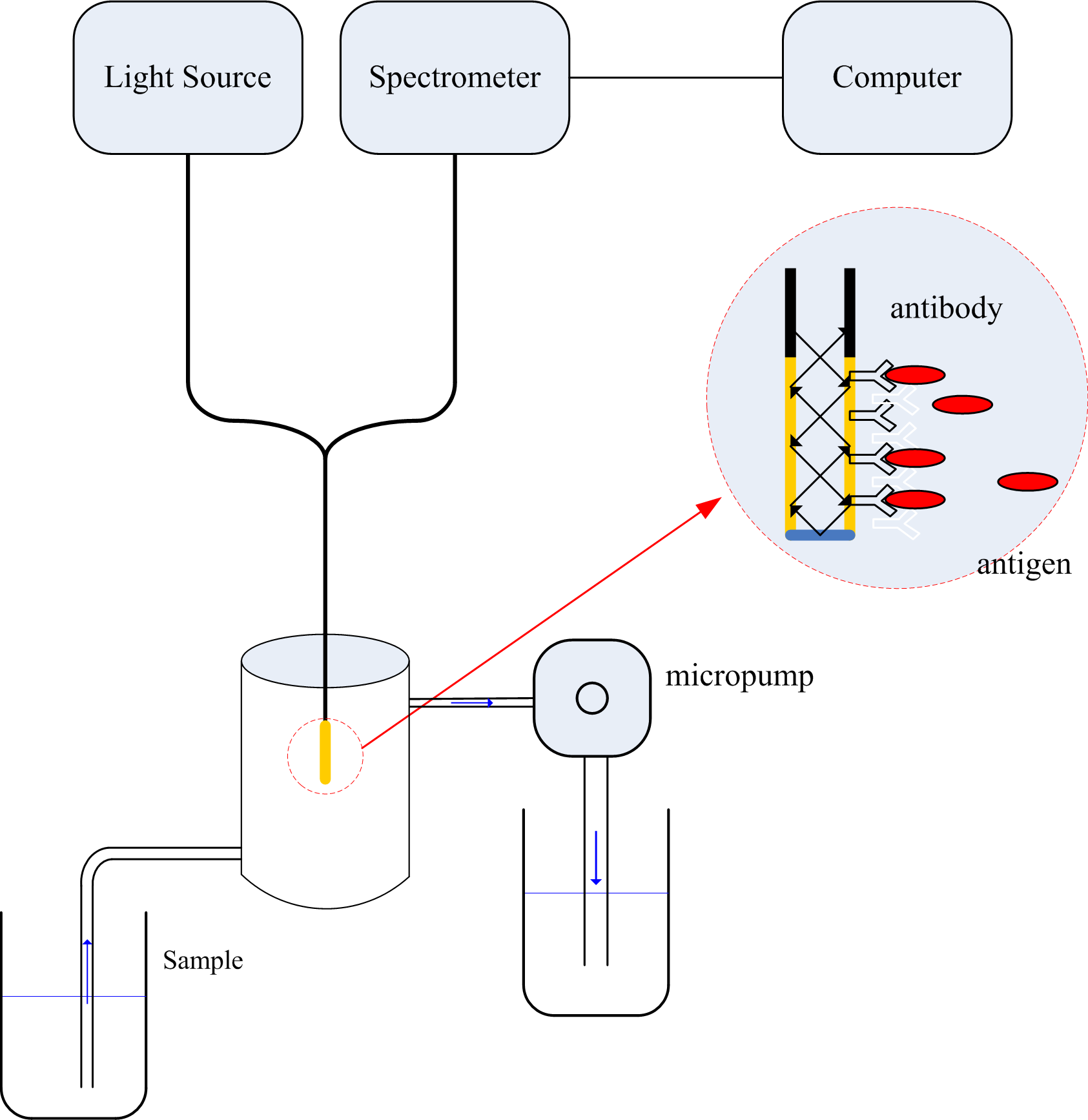}
  }
  \subfloat[]{ \raisebox{10mm}{
    \includegraphics[width=0.4\textwidth]{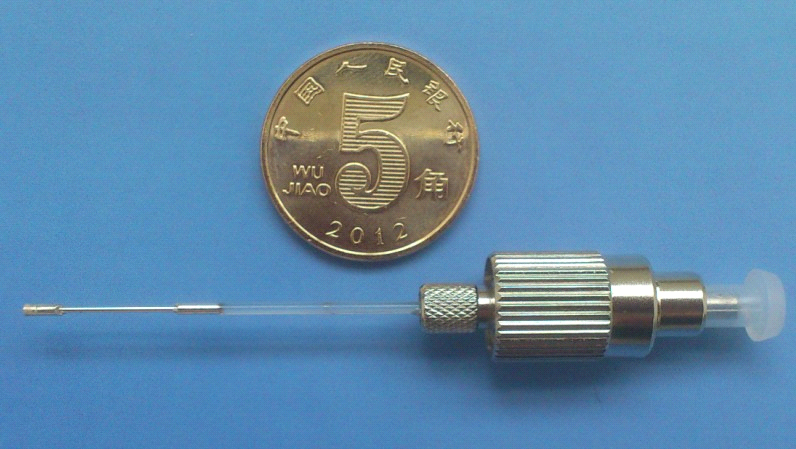}      
    \label{fig:probe}
  }}

  \caption{(a) System configuration of the fiber SPR measurement
    system. A model of affinity interaction for a biochemical
    experiment is presented in the red circle. (b) Fiber SPR sensor
    based on multi-mode fiber with silver film.}
\end{figure}

\subsection{ Experimental configuration and measurement}

The experimental configuration is shown in
Fig.~\ref{fig:schematic}. The fiber optical SPR system comprise of a
fiber SPR sensor and a measurement system. The fiber sensor is
fabricated from a multi-mode fiber with numerical aperture 0.39
produced by Polymicro Inc. The diameter is $600~\mu \mathrm{m}$ coated
with plastic cladding.  This plastic coating can be easy strip down
and cleaned with acetone solution. The metal film is deposited with an
ion sputtering vapor system. This device is a common one for plane
target and the chamber vacuum is below $8.5\times 10^{-3}$Pa. We
designed a gear commutator which rotates the fiber above the
sputtering disk. Therefore, we achieve a uniform coating on the fiber
cylindrical surface. The thickness of the deposit film is controlled
by the exposed time. Since it is a round fiber, we cannot have an
accuracy thickness measurement directly. The thickness of silver film
is estimated about $45~\mathrm{nm}$. Later a thick film is deposited
at the fiber end which acts as a reflection mirror. The red circle if
Fig~\ref{fig:schematic} present a model of affinity interaction for
future development.

The light source is an Oceanoptics DH-2000-BAL with high luminescent
tungsten halogen bulb. The spectrum is from 450 nm to 1100 nm, shown
in Fig.~\ref{fig:gui} below. The light is delivered to the sensing
region by a $600~\mu \mathrm{m}$ combiner and the transmission
spectrum is measured by a spectrometer. The spectrometer is a portable
model USB4000 from Oceanoptics. This spectrometer has 3648 pixels with
measurement range from 345 nm to 1043 nm, cover the surface plasmon
resonance region of interesting. Spectral data are collected to the
computer through USB port and the transmission time can be ignored.

\begin{figure}[hbt]
  \centering
  \includegraphics[width=0.7\textwidth]{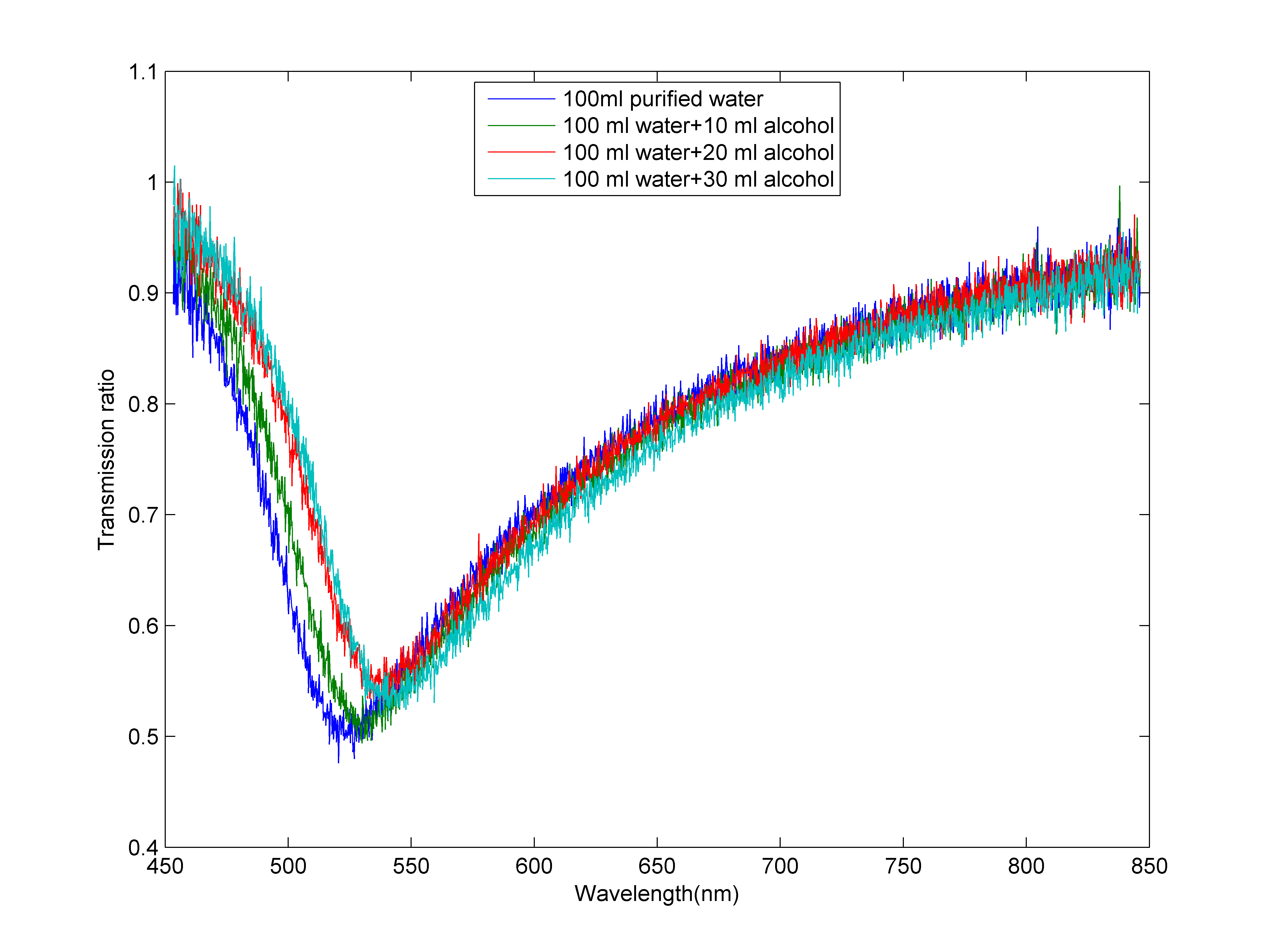}
  \caption{SPR curves of purified water and water alcohol mixture.}
  \label{fig:groupspr}
\end{figure}

Fig.~\ref{fig:groupspr} shows a group of experimental results with
water and ethanol mixture of different concentration. The transmission
ratio is similar and greater than 50\%. While the ratio of alcohol
increases, the refractive index of solution increase and the resonance
wavelength move toward the long wavelengths.

\subsection{Integrated measurement environment}
Based on the system hardware above we can perform a basic fiber SPR
measurement with the Spectrasuite software developed by Oceanoptics
Inc. Since Spectrasuite is developed for general-purpose usage, and
not dedicate for SPR measurement with real-time response, the
measurement process is cumbersome. Thus we developed an integrated
measurement environment with real-time spectrum display and result
show. We prefer java as the development language. The primary reason
is that the spectrometer has a native java driver and there are many
mature libraries which may help the software development. Java is also
well known for its cross platform ability. This may help in the
software mitigation for the potential industrial application.

\begin{figure}[tbh]
  \centering
  \subfloat[]{
    \includegraphics[width=0.55\textwidth]{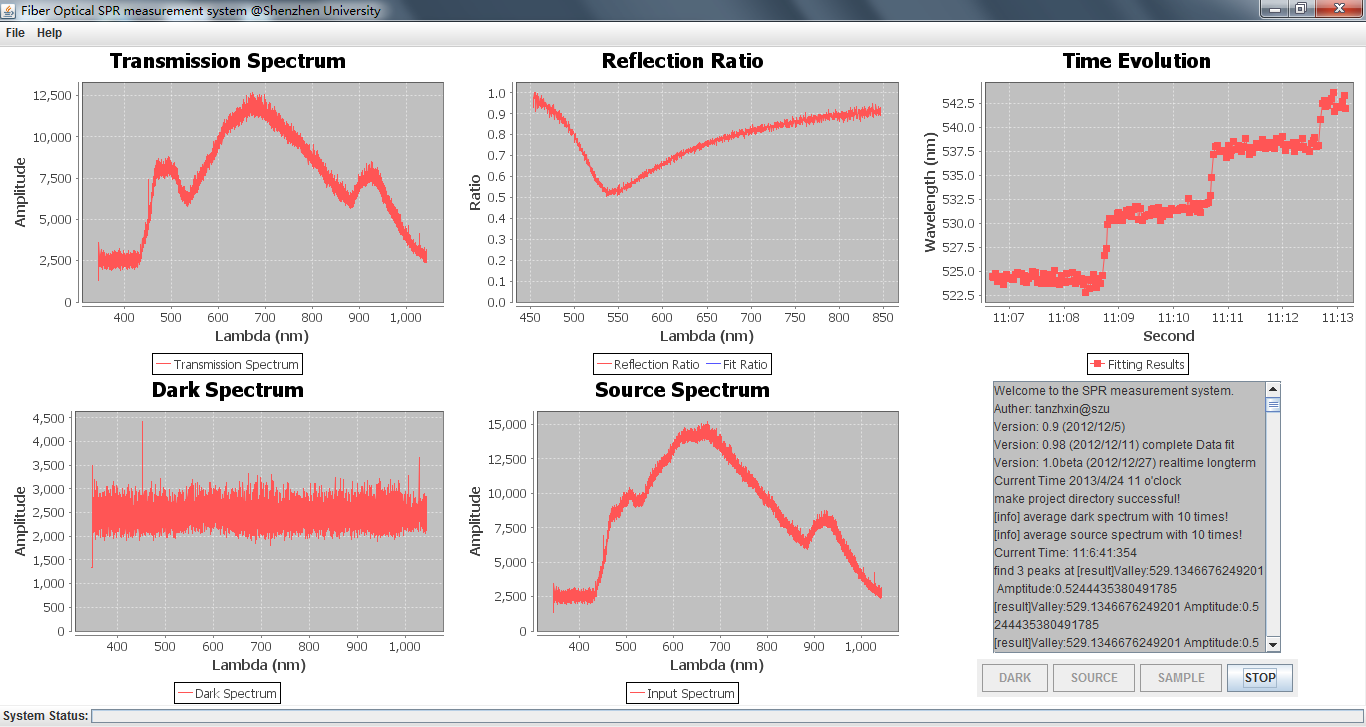}
    \label{fig:gui}
  } \subfloat[]{
    \includegraphics[width=0.38\textwidth]{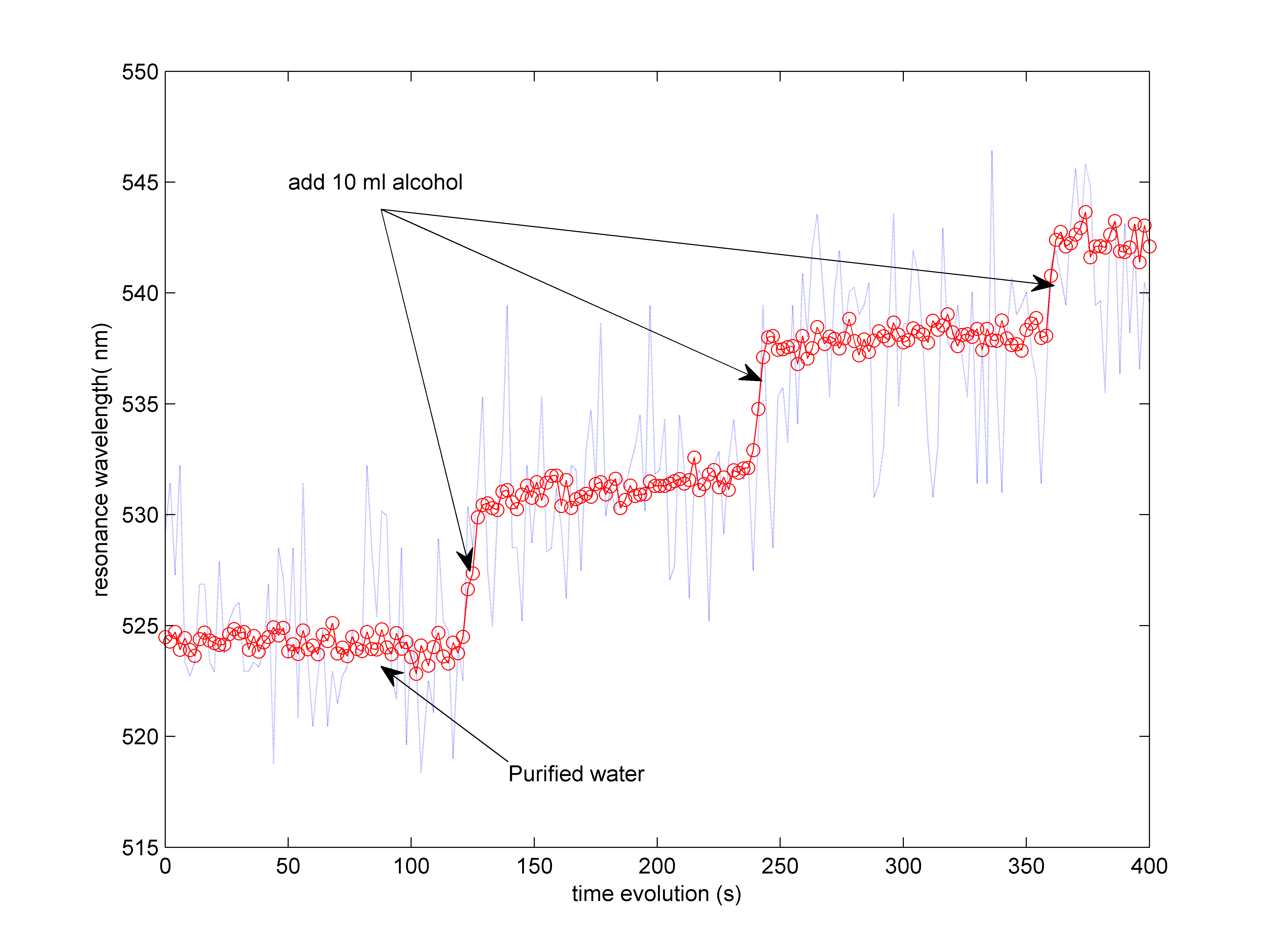}
    \label{fig:fitted}
  }

  \caption{(a)~Snapshot of the integrated measurement environment. The
    upper right panel presents the evolution of the resonance
    wavelength for biochemical affinity analysis. (b)~Resonance
    wavelength before and after data fitting process. }
\end{figure}

The software interface is shown in Fig.~\ref{fig:gui}. The main panel
is divided into six zones, including a control panel and five spectrum
displays. There are a dark spectrum, a source spectrum, a transmission
spectrum, a reflection ratio display and the evolution of the
resonance wavelength. The dark spectrum and the source spectrum are in
the bottom row. The dark spectrum is the background electronic noise
from the liner CCD of spectrometer and environmental light. Whenever
the fiber optical SPR sensor is immersed in the analyte solution, the
transmission spectrum is measured repeatedly, shown in the upper left
corner. The upper center display is the transmission ratio with
definition.

\begin{equation}
  \mathrm{R} = \frac{\mathrm{I_{trans}} - \mathrm{I_{dark}}}{\mathrm{I_{source}}-\mathrm{I_{dark}} }
  \label{eq:ratio}
\end{equation}

The original spectra are truncated to eliminate the numerical
instability caused by the small light intensity in
Eq.~\ref{eq:ratio}. The reflection ratio chart (upper center panel) is
a typical fiber SPR curve with a broad FWHM diagram, and the
reflection ratios are greater than 0.5. The blue line is the data
fitting results and we will discuss below. The chart in the upper
right corner is the time-evolution of the sensing interaction. The red
points represent the resonance wavelength after the data fitting
procedure. In this chart the resonance wavelength changed in step
shape as we add alcohol to the purified water. The control panel is in
the bottom right which contains the control buttons and the output
information.

As shown in Fig~\ref{fig:gui}, with this software we achieve an online
measurement with optical fiber sensor which is remarkable for a
biochemical interaction analysis. In the software development, the
main issues that may affect our measurement are the real-time response
ability and the data fitting process, and we discuss below.

\subsection{Real-time response ability}

The data acquisition involves a setting of CCD integration time which
should adapt to the measurement. The overall integration time should
small than that of the timer setting which is determined by the
updating frequency. The real-time response may be slowed by the
graphical toolkit and the data fitting procedure if we handle it
carelessly. The graphical user interface is based on JFreechart which
is a famous free Java chart library released under the terms of the
GNU LGPL. It is a rich-featured chart library. Since whenever the
chart dataset is updated, the program should repaint these charts
completely. This may be limited by the ``frames per second" rate that
we can achieve with the graphical library. JFreechart can afford a
refreshing speed of once per second and updating multiple times per
second may results in high CPU load. Actually, the data fitting
process is very quick since it is a matured algorithm.

In common a biochemical interaction will last for several minutes to
tens minutes.  Therefore we may update the resonance wavelength every
second. Consequently, this will get over hundreds data points which is
enough for the data process. Moreover, JFreechart is on developing
which may improve its real-time response ability in the future. So,
this graphical library is enough to satisfy the real-time requirement
for SPR analysis.

\subsection{Data fitting for the resonance wavelength}

Since the fiber optical SPR sensor based on multi-mode fiber is a
mixture system with light in various wavelengths and incident in
different angles, the transmission ratio is quite different from a
simple Gaussian shape. Therefore, we prefer a general polynomial
fitting scheme. The data region is around the valley position with a
length of 300 pixels, about 70 nm. Then we recalculate the value at
each pixel and get the absorption peak position. The fitting procedure
shows quite well result in a 95\% confidence and the fitting curve
keeps the data shape and smooth out the fluctuation.

Fig.~\ref{fig:fitted} shows the wavelengths before and after the data
fitting procedure. The dotted blue line represents the minimum
position among the ratio which is from a peak find function provided
by Oceanoptics. The red points are the resonance wavelengths fitted
with polynomial function.  Since data fitting is a key issue in our
sensor and it provides the ultimate resonance wavelength for the
biochemical interaction analysis, we would like have a detail
discussion about our experience. We have tried the average
method. Multi-time average method will decrease the real-time response
ability. Meanwhile, the fluctuation in reflection ratio chart cannot
be suppressed by average because of the intrinsic properties of the
spectrometer. For a common measurement, we can average two results to
reduce the systematic error. The data fitting procedure may
incorporate an interpolation algorithm to increase to resolution of
the sensor. The premise is that the fiber sensor system should have a
good dynamic resolution than that of the spectrometer. The resolution
can be estimate by the variation of the time series of resonance
wavelength which correspond to the minimal RI change that the fiber
sensor can measure. The variation of the resonance wavelength may be
influenced by many factors, such as film thickness, film pattern, and
light source stability et al. Currently, the micro-fluid system is not
implemented and the experimental measurement are conduct in simple
configuration which increase the experimental variation.

\subsection{Improvement in the future}
Since the intensity of source light may decrease in a long-term
monitor and the spectrum changes, we may add another spectrometer to
synchronously measure the spectrum of light source. The fluidic system
is in our plan. Another issue is to improve the system resolution with
interpolation and algorithm improvement.


\begin{figure}[bth]
  \centering
  \includegraphics[width=0.7\textwidth]{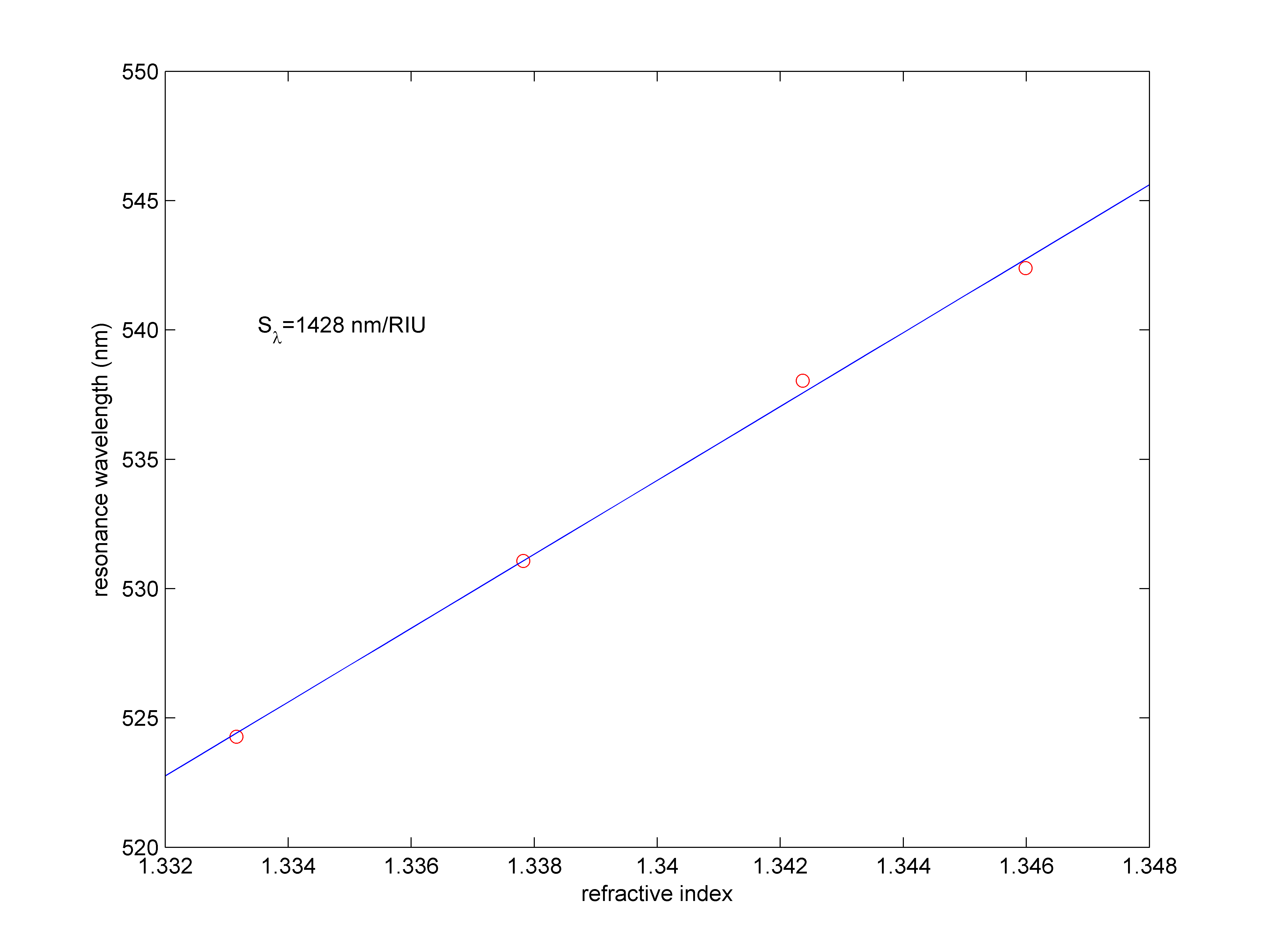}
  \caption{The sensitivity of the fiber sensor}
  \label{fig:sensitivity}
\end{figure}

\section{Application for chemical and biochemical interaction}

Fig.~\ref{fig:sensitivity} shows the processing result of the dynamic
measurement. The resonance wavelengths are averaged with data in the
middle of step region.  The refractive index of mixture is
interpolated based on the volume ratio of
alcohol~\cite{Nowakowska1939RefractiveIndicesEthyl}. The sensitivity
of fiber sensor is 1428 nm/RIU. Therefore, a corresponding resolution
in RI is about $1\times 10^{-4}$ RIU. For a typical biochemical
interaction with 0.002 RI changes, the data interval in y-axis is near
20 which are decent for a data processing.  In
Ref.~[\citenum{Yanase2010Developmentopticalfiber}] the authors had
made a measure of bio-chemical interaction with the fiber SPR sensor
in a similar configuration. Although the resolution of fiber SPR
sensor is much low than that of a prism configuration, this gap may be
decreased or filled by some special design, such as fiber SPR sensor
with photonic crystal
fiber~\cite{Hassani2007Designcriteriamicrostructured,Popescu2012Powerabsorptionefficiency}. The
characteristic of fiber optical SPR sensor with in situ measurement
and compactness may meet some special application, for example, a fast
detection in vivo~\cite{Kim2012Invivooptical}.

\section{Conclusion}
We construct a fiber optical SPR system and build an integrated
measurement environment to record a real-time response for biochemical
interaction analysis. The fiber optical SPR sensor is measured with
water and alcohol mixtures of different concentration. The system
real-time performance and data processing are discussed in
detail. This integrated fiber SPR measure system will help to realize
a minimized SPR biosensor with in situ and real-time measurement
ability.

\section{Acknowledgement}
This work is supported by the National Science Foundations of China
under Grants (No.61275125), the Research Foundation for the Doctoral
Program of Higher Education of Ministry of Education (20124408110003),
the Province-Ministry Industry-University-Institute Cooperation
Project of Guangdong Province under Grant (No. 2010B090400328), the
Shenzhen Science and Technology Project, the Shenzhen Nanshan District
Science and Technology Project, the High-level Talents Project of
Guangdong Province.


\end{document}